\documentclass{elsart}
\usepackage{epsfig}

\begin{document}


\begin{frontmatter}

\title{The DELPHI Silicon Tracker in \\ the global pattern recognition}

\author{M. Elsing}

\address{EP Division, CERN, CH-1211 Geneva 23. Switzerland}

\begin{abstract}
\noindent
ALEPH and DELPHI were the first experiments operating a
silicon vertex detector at LEP.
During the past 10 years of data taking the DELPHI Silicon Tracker was upgraded
three times to follow the different tracking requirements for LEP 1 and LEP 2
as well as to improve the tracking performance. Several steps in the
development of the pattern recognition software were done in order to
understand and fully exploit the silicon tracker information. This article
gives an overview of the final algorithms and concepts of the
track reconstruction using the Silicon Tracker in DELPHI.
\end{abstract}

\end{frontmatter}


\section{Introduction}

Since 1990 DELPHI has been operating a Silicon Tracker
\cite{VD-first} in the barrel region close to the beam pipe. This
detector has been upgraded \cite{VD-23} three times to follow the different
tracking requirements for LEP 1 and LEP 2 as well as to improve the tracking
performance. In its final upgraded version the Silicon Tracker
\cite{VD-LEP2} covers nearly the full polar angle down to $10.5^\circ$.
It is the innermost detector of the tracking system of the
DELPHI detector \cite{DELPHI}, which is one of the most complex because of
the presence of the Ring Imaging Cherenkov Counters (RICH) in the central
(barrel) and the endcap (forward) regions. 

The global track reconstruction software
was developed in parallel to the upgrades of the Silicon Tracker.
First versions of a pattern recognition were adding the Silicon Tracker
hits to tracks which were reconstructed using the outer tracking detectors.
In 1995 the
so called "$R_b$ crisis" \cite{Rb-crisis} led people to carefully study
reconstruction problems which were limiting the $b$-tagging performance. 
It was realised that the
method of using the Silicon Tracker information at the end of the
reconstruction was not sufficient for a
precise $b$-tagging and for an efficient vertex reconstruction in $\tau$ and
heavy flavour decays. New track reconstruction algorithms starting from the
Silicon Tracker information were needed to solve the problems.
The development of a new reconstruction package ended in 1999
with the complete integration of the forward part of the Silicon Tracker
(Very Forward Tracker, VFT) into the global pattern recognition.

This article is structured as follows. At the beginning a brief introduction
to the Silicon Tracker and the DELPHI outer tracking system is given. Then the
track reconstruction concepts and algorithms are discussed in the two following
sections. In the third section a description of the
precision tracking in the barrel part
is given starting from the reconstructed Silicon Tracker clusters and the
measured track elements in the outer tracking detectors. In the fourth section
the use of the VFT hits in the forward tracking is discussed.
The performance of the track reconstruction software in the barrel and forward regions
is shown.

\section{The layout of the Silicon Tracker and the DELPHI outer tracking
system}

In the DELPHI standard coordinate system the $z$ axis is along the electron
direction, the $x$ axis points towards the centre of LEP and the $y$ axis
points upwards. The polar angle to the $z$ axis is called $\theta$ and the
azimuthal angle around the $z$ axis is called $\phi$. The radial coordinate is
$R = \sqrt{x^2+y^2}$.

\begin{figure}[htb]
 \begin{center}
  \epsfig{file=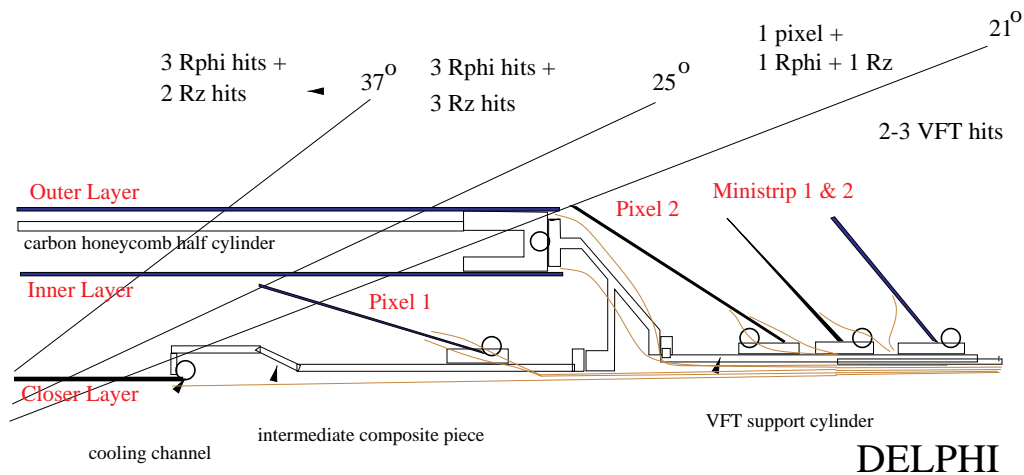,width=13.5cm}
 \end{center}
 \caption[]{\label{VDfig} A cross section of a quarter of the DELPHI Silicon
Tracker for $z > 10~c$m. See text for details.}
\end{figure}

Figure \ref{VDfig} shows an $Rz$ cross section
of a quarter of the DELPHI Silicon Tracker. A detailed description of the
detector (in its final setup) can be found in \cite{VD-LEP2}.
It is divided into a barrel part (Vertex Detector, VD) and the VFT in
the forward direction. For the installation around the beam pipe the
mechanical structure is divided into two half shells in $R\phi$.

\begin{figure}[htb]
 \vspace{-7cm}
  \mbox{\hspace{-2.cm}\epsfig{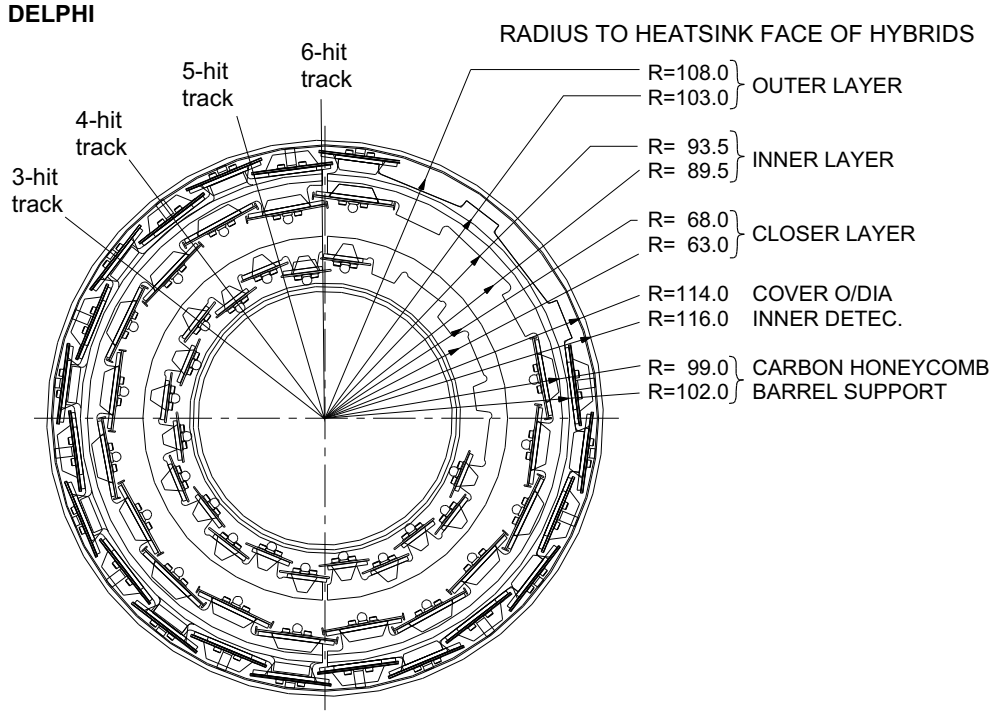}}
 \vspace{1.25cm}
 \caption[]{\label{VDrphi} A $R\phi$ view of the barrel part of the Silicon
Tracker showing the Outer, Inner and Closer layer. The hybrids are
overlaid to the sensors.}
\end{figure}

The VD consists of 3 concentric layers (called Closer, Inner and Outer)
at average radii of 6.6 cm, 9.2 cm and 10.6 cm, respectively. All three layers
cover polar angles of $25^\circ-155^\circ$,
the Inner layer extends the coverage to $21^\circ - 159^\circ$.
The Closer as well as the Outer layer is made of 24 modules with a 15 \%
azimuthal overlap, while only 20 modules are used for the Inner layer
(figure \ref{VDrphi}). Each module in the Outer and the Inner layer consists
of 8 sensors, the Closer layer modules are shorter and consist of only
4 sensors. Half of each module is bonded in series and
read out at the outer ends. All Closer and Outer layer
modules have
double sided readout to measure $Rz$ and $R\phi$, while only
the "extreme" 2 sensors in the Inner layer are double sided, the "central"
sensors measure only $R\phi$.
The n-side lines of one sensor of the "flipped" modules in the Closer and the Inner
layer are connected to the p-side ones of the adjacent sensor.

The hit resolution in $R\phi$ is $\sim 8 ~\mu$m. In the $Rz$ plane
the readout pitch is changed for plaquettes at different angles to give the
best resolution perpendicular to the track, varying between $\sim 10 ~\mu$m
and $25 ~\mu$m for tracks at different inclinations.

The VFT consists of two pixel layers, the first one being located inside the
VD, and two mini strip layers. It covers the angular region of
$11^\circ - 26^\circ$ and $154^\circ - 169^\circ$. The pixel dimension
is $330 \times 330 ~\mu$m$^2$. The two layers of back-to-back mini strip detectors
have a readout pitch of $200 ~\mu$m and one intermediate strip. To help the
pattern recognition the mini strip modules are mounted at a small stereo
angle.

\begin{figure}[t]
 \begin{center}
  \epsfig{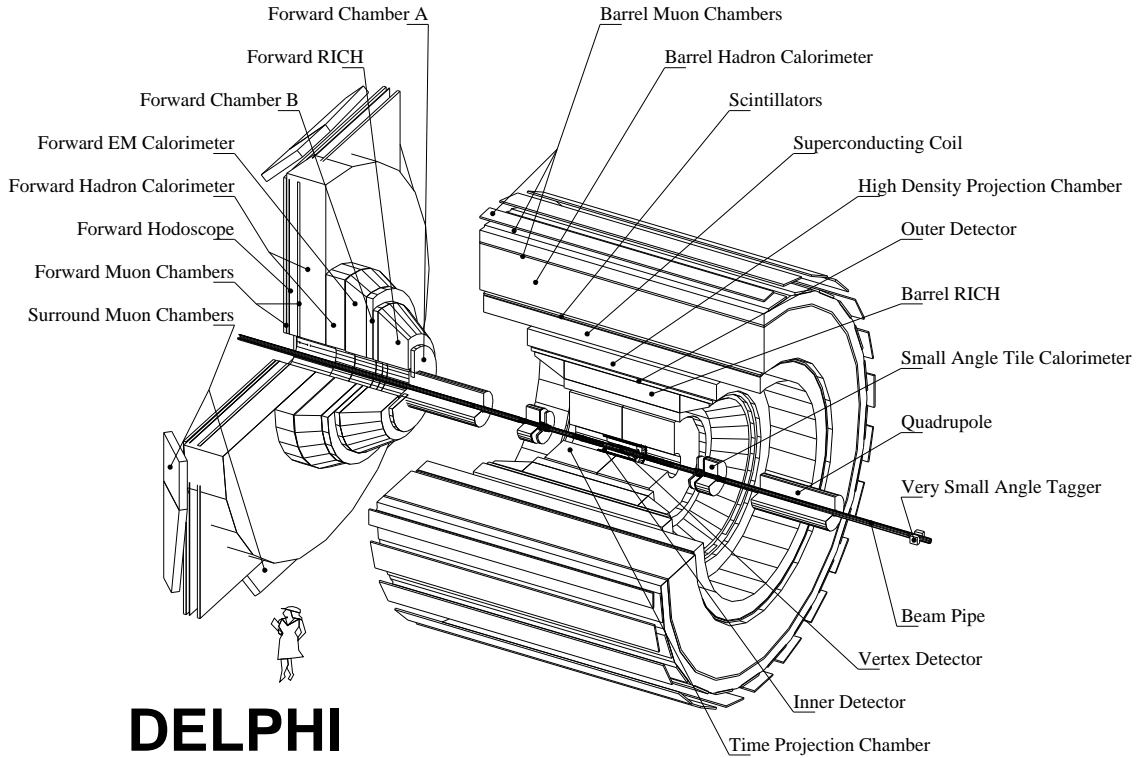}
 \end{center}
 \vspace{1cm}
 \caption[]{\label{DELPHI} A schematic view of the DELPHI detector.
The tracking system can be divided into a barrel and a forward region
following the overall detector layout of the barrel and the endcap parts.
See text for details.}
\end{figure}

A detailed description of the DELPHI apparatus can be found in
\cite{DELPHI}. A schematic view of the detector
is shown in figure \ref{DELPHI}. The DELPHI outer tracking system is
divided into a barrel and a forward part.

In the barrel part precise tracking information is
provided by the Time Projection Chamber (TPC) and the
Inner (ID) and Outer (OD) Detectors. The ID consists of a jet chamber 
to perform a precise $R\phi$ measurement and of 5 concentric layers
of straw tubes. It is followed by the TPC, which covers polar angles
between $21^\circ$ and $159^\circ$. The TPC single point resolution for
charged particles is $250 ~\mu$m in $R\phi$ and $880 ~\mu$m in $z$. The
Outer Detector is mounted behind the Barrel RICH to give additional tracking
information at a radius of $2.02$ m. The five layers of drift cells cover polar
angles between $42^\circ$ and $138^\circ$ and provide $R\phi$ and $z$
information.

In the forward region the tracking is improved
by two planar drift chambers in front of and behind
the Forward RICH. Their polar angle coverage is
$11^\circ-33^\circ$ (FCA) and $11^\circ-36.5^\circ$ (FCB), respectively.
The Forward RICH provides an additional track point from the ionisation
of the charged particle inside the drift box.

\section{Precision tracking in the barrel part}

The detectors of the outer tracking system in the barrel region
measure the tracks of charged particles with high redundancy.
Local pattern recognition algorithms are used to reconstruct the track
elements in the different detectors.
The VD cluster reconstruction algorithm is documented in \cite{Ken}.
A careful treatment of the VD
hit information and good simulation description are needed together with
an optimised track reconstruction software in
order to exploit the high precision tracking information in the barrel region.

\subsection{Internal VD alignment}

The internal alignment of the Silicon Tracker is based on a mechanical survey,
during which the components and the whole structure were measured by optical or
mechanical means. The survey was done in two steps \cite{survey}. First a
measurement of each individual module was done to fix the position of a
sensor within a module  to a level of 1-2 $\mu$m. Then the position of each
module in the two half shells was determined with a relative precision of
about 10 $\mu$m.

The detector may significantly deform after the
survey during transportation and installation. Therefore
the final precision is obtained by an \mbox{offline} alignment
using tracks from charged particles.
The first step after the installation of the detector
is the determination of the relative position of the half shells w.r.t. each
other using charged tracks crossing the top or bottom overlaps of
the two halves.

Several effects which influence the detector alignment are taken into
account in the alignment procedure \cite{VChabaud}. The Lorentz angle effect
leads to a shift of $\sim ~6 ~\mu$m in $R\phi$ ($B=1.2$ Tesla). Due to the flipped
module design the resulting shift for a reconstructed hit is opposite for
sensors where the p-side is facing the beam pipe or the outer tracker. Moreover, in the
study of the Lorentz angle effect it has been found
out that the barycentre of the holes and electrons created by the charged
particle passing through the detector
does not correspond exactly to the mid-plane of the detector.
A 10-20 $\mu$m shift of the barycentre of holes and electrons in the radial
position towards the p-side needs to be taken into account.
Finally bowing of individual modules along $z$ of up to 150 $\mu$m in radius
has been observed. This effect is related to
stress during installation and the size of the bowing varies with temperature
and humidity.

The final alignment \cite{align} is done using charged tracks
in $Z \to \mu\mu$ and hadronic events. The Outer layer
(see figure \ref{VDrphi}) is taken as a master layer for
the whole detector. Its modules are aligned w.r.t. each other using hadronic
tracks passing the 15 \% azimuthal overlap between adjacent modules. The Closer
layer is aligned w.r.t. the Outer layer using muons in $Z \to \mu\mu$
events, where the momentum of each muon track is fixed to the nominal
value. The Inner layer is aligned to the other two using again
tracks from hadronic events. This procedure is iterated a few times until a
stable alignment is found.
At the end an overall twist of the detector around the beam
axis is measured \cite{align} using the geometrically signed impact parameter 
of tracks w.r.t. the beam spot as a function of their polar angle $\theta$.

\subsection{Shaken VD alignment for the simulation}

The intrinsic $R\phi$ hit resolution of a VD sensor is 5 $\mu$m,
while in DELPHI a resolution of $\sim 8 ~\mu$m
\cite{VD-LEP2} has been achieved. The difference reflects imperfections in
the internal alignment and small deformations in the flatness of the sensors.

The simulation needs to reflect the actual precision obtained for the data. For
example a simple hit smearing would not allow for correlations between
tracks, because in real data tracks hitting the same
sensors are affected in the same way by residual problems.
Therefore a scheme of shaking the detector position in the reconstruction
of simulated events
is used to model the actual resolution. Such a scheme also
includes effects of misalignment at the track
reconstruction level into the simulation.
The position of each sensor
is varied from its nominal position on an event by event basis \cite{Ken}.
This is needed in order not to have any visible pattern in the misalignment
of the simulation. Such
a pattern would have been cured by the alignment procedure on the real data.
The RMS of the shifts applied were e.g. 6.4 $\mu$m in $R\phi$, 7.3 $\mu$m
in $z$ and
20 to 37 $\mu$m in $R$ for the simulation corresponding to the 1995 real data.
Typical rotations are of the order of 0.15 - 0.4 mrad.

\subsection{Quality cuts on cluster signal over noise}

It is necessary to remove noise clusters from the event to avoid tails in
the impact parameter resolution function.
Cuts on cluster signal over noise are applied to
improve the purity of good hits
without losing much efficiency. The VD
is made of several different sensors which have different
signal over noise performance. Typical values range from 10 to 18 and close to
30 for the p-side of the single metal Outer layer \cite{VD-LEP2}.
Hence the cuts on signal over noise are tuned for each module and vary from
6 to 15. The cuts are applied to
the hits at association time and depend on the track topology to
improve the efficiency.

Tracks having $R\phi$ hits in 3 out of 3 layers
are very tightly constrained and therefore the chance of picking up a random
noise hit is very small. For such tracks it is required that only 2 out of 3
associated $R\phi$ hits pass the signal over noise cut.

\subsection{Efficiency correction for the Simulation}

The description of the hit efficiency of the detector
is an important aspect of the simulation.
Dead modules and dead channels need to be taken into account too. A sensor by
sensor tuning \cite{Ken} of the hit efficiency is done in the simulation by
randomly removing hits from the event in order to match the apparent
efficiency to the one in the real data.
Modules which are inefficient for only parts of the data taking period are
allowed for by dropping
the module information from a corresponding fraction of the
simulated events.

\subsection{Kalman Filter track fit}

The task of the track fit is to determine the track parameters from the
combination of VD hits and track elements measured in the outer tracking
system. The track fit algorithm
used by DELPHI is a Kalman Filter \cite{trackfit}, which is a fast 
recursive algorithm. It is implemented using the weight matrix
approach.

A Kalman Filter is an estimator for a linear system, while a track
in a solenoid field is described by a helix. A Taylor expansion around
the reference trajectory is used as a starting point to obtain a linear system.
The fit is iterated to ensure good convergence.

The DELPHI track fit takes into account the effects of multiple scattering
and energy loss of particles in the material.
A simplified description of the detector material is sufficient for the
purpose of track fitting. The geometry of the detector material of the outer tracker
is approximated by a sequence of surfaces, which are either cylinders around or
planes perpendicular to the beam pipe.
For each of these surfaces an apparent thickness is specified in terms of
radiation length and energy loss of a minimal ionising particle (for a particle
crossing at $90^\circ$).

A different approach is used for the VD material in the fit.
Multiple scattering and energy loss in the material of the VD and the beam
pipe are dominant contributions to the impact parameter resolution.
Therefore a more detailed description of the material distribution is used
for the track extrapolation and fitting inside the VD. The
description reflects the
complicated support structures and the overlap of modules in the individual 
layers in the corresponding azimuthal regions as can be seen
in figure \ref{VDrphi}.

In the track fit the effect of the multiple scattering is taken
into account by increasing the error contour of the track extrapolation after
crossing the material surface. The momentum dependent effect of the energy
loss is taken into account by changing the curvature of the
reference trajectory used for the Taylor expansion in the fit.

Another important feature of the DELPHI fit is the logic to remove
outliers. The fit is able to remove up to 3 measurements from a
track candidate if it fails a fit $\chi^2$ probability cut of 0.1 \%.
This is a very effective filter to remove wrong
associations of hits to tracks. A ranking of detectors to be removed is
used in order not to remove the most precise measurement (e.g. the TPC
track element) from the track. Called from the track search packages the fit
always retains the track element which was used as a starting point to
reconstruct the track.

\subsection{Optimised track search algorithms}

The first version of a DELPHI track search \cite{TPCpivot} was
based on the track elements found in the TPC. These track elements were
extrapolated to the ID jet chamber and the OD to associate additional hits.
The output of the searches was a sample of
candidates. The track parameters for all of
the candidates were determined by the track fit and bad combinations were
removed. Remaining ambiguous associations were resolved by a two stage
process selecting good tracks. No VD information was used to further
constrain the ambiguity decision.
The VD hits were associated afterwards.
In a first step all tracks were extrapolated to
the VD layers and the $R\phi$ hits were associated track by track, in
the second step the $Rz$ hits were associated. After the association
all tracks were fitted to include the VD information
in the track parameters. Any mistake done in the linear chain of reconstruction
steps resulted in a problem for the following steps. This led to rather
unstable results which furthermore were not reproduced in the simulation.
The performance of the package was strongly dependent on the track density
and therefore problems were more frequent in events with heavy quarks.

The new track reconstruction software
is using a completely
different approach. Now the VD is used as a starting point for the
track search, because it is the most precise detector and has the best two track
resolution of all tracking detectors in DELPHI. There are two new track
search algorithms, one is \mbox{using} all combinations of TPC tracks and
2 or 3 associated $R\phi$ hits in the VD \cite{TPCVD}, the other
is starting with ID jet chamber hits plus VD $R\phi$ hits \cite{IDVD}.
The algorithms were designed in $R\phi$ only without using VD
$Rz$ hit information, because the Silicon Tracker had no double sided
readout for half of the LEP 1 data taking period. In both new searches
bad combinations are filtered using the track fit. Remaining candidates are
extrapolated to the other detectors to search for possible
associations of additional hits. Each association is tested again using the
track fit outlier logic. All ambiguous track combinations are then fed into
a new ambiguity processor to resolve the full event. This processor will be
discussed later in this paper.

The new reconstruction software uses the VD $Rz$ hits, which are present for
the last two years of LEP 1 data taking and for the complete LEP 2 data set,
to improve the resolution.
The $Rz$ hits are associated to the resolved tracks
found using only the $R\phi$ hit information.
All possible associations of $Rz$ hits to tracks are considered and
filtered using the full
track fit. The ambiguities are then resolved in a second run of the global
event ambiguity processor. 

\subsubsection{Secondary interactions in material}

\begin{figure}[htb]
 \vspace{3cm}
 \begin{center}
  \epsfig{file=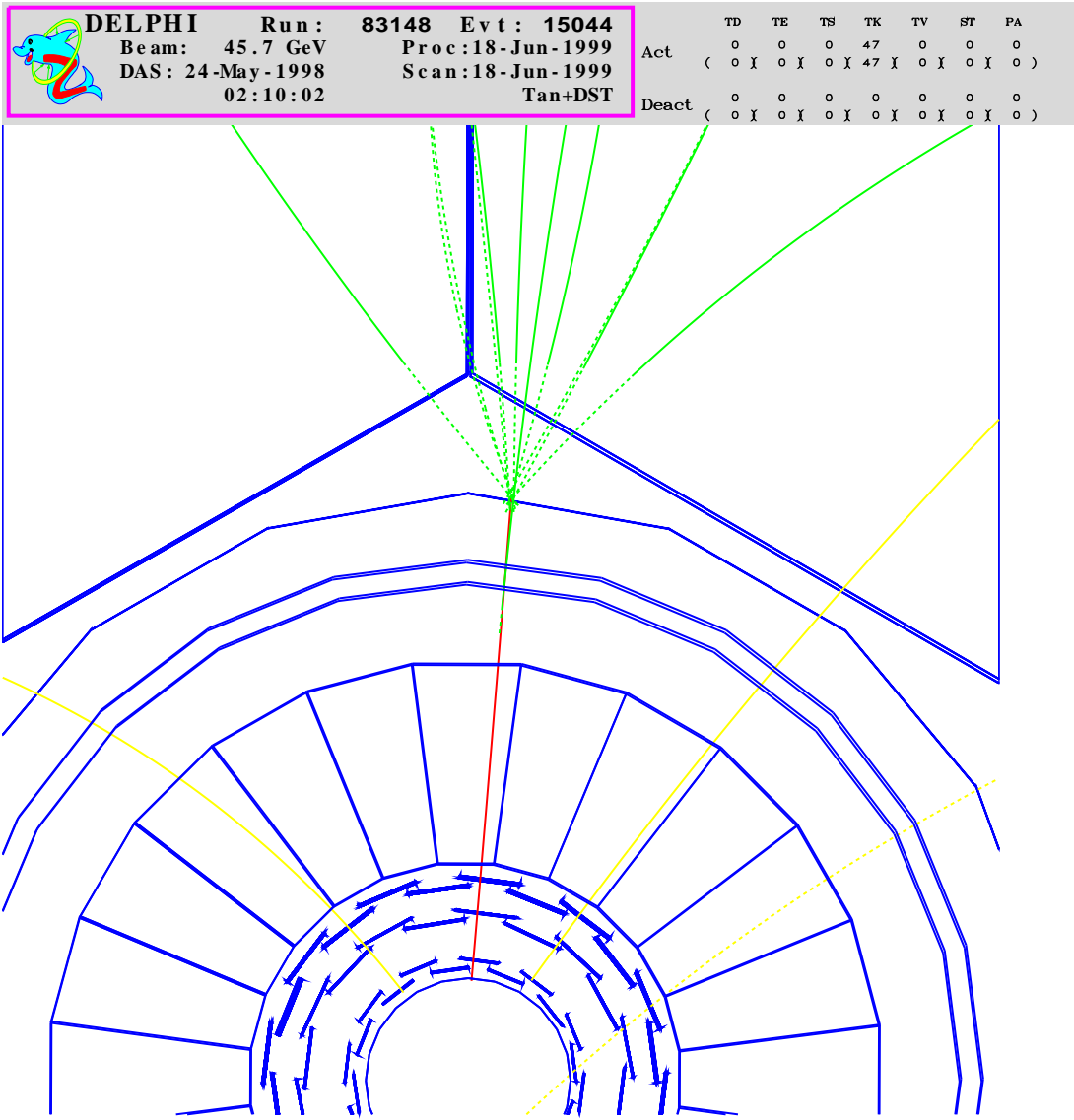,width=15cm}
 \end{center}
 \vspace{-4cm}
 \caption[]{\label{inter} A reconstructed hadronic interaction in the material
in front of the TPC. Shown is the central part of the DELPHI detector with the
VD layers at the bottom. The track before the interaction is reconstructed
using the VD and the ID jet chamber.}
\end{figure}

Searches for vertices from hadronic interactions, from $\gamma$
conversions and from decays of $K^0_s$ and $\Lambda$
are part of the new tracking software \cite{mammoth}. 
Figure \ref{inter} shows an example of a hadronic interaction in the detector
material in front of the TPC.
The TPC track elements of secondary particles produced due to
such interactions or decays are a problem for the
track reconstruction, because wrong association of VD hits to secondary tracks
would disturb the correct association of VD hits to primary tracks.

Therefore the vertex searches are called before the track searches to
reconstruct such secondary vertices using only the TPC track elements. All
track elements associated to the secondary vertices are removed from
the event before the full reconstruction of primary tracks. No
VD hits are associated to these TPC
track elements. A dedicated track search is then performed to reconstruct
the tracks which are pointing to the hadronic
interaction vertices \cite{IDVD,IDVDonly}. This search uses the remaining
unassociated hits in the VD and the ID jet chamber after the reconstruction of
primary tracks which did not cause a hadronic shower before the TPC.
The tracks measured only in the VD or the VD and the ID jet chamber are then
linked \cite{mammoth} to the secondary vertices
to fully reconstruct the track-vertex structure. Finally elastic
interactions and decays in flight of pions or kaons are reconstructed.

\subsection{The concept of "exclusions"}

The result of track searches is a set of ambiguous track candidates with 
many possible associations of individual hits to different track candidates.
Also the local pattern recognition of the different detectors
can create ambiguous hit combinations.
For example in the VFT mini strip detector
space points are reconstructed out of hits on the back-to-back module by
combining the measurements in both orientations so that
hits from $n$ tracks on a module lead to $n(n-1)$ mirror images.
In the ID jet chamber the left/right ambiguity can not be
resolved using only the detector information itself.
All such ambiguities need to be resolved
in the process of the track reconstruction. It is beneficial to leave the
decisions to the stage of the global event solution, since at this stage
the full information of the reconstructed track candidates can be used to
minimise mistakes.

All results from the
different reconstruction packages are stored in the
DELPHI event database \cite{tanagra}.
The database structure allows for so called "logical exclusions" between
objects like hits or track candidates. An "exclusion" signals that two objects
use conflicting or common detector information and that for the final
solution of the event such conflicts need to be resolved.

\subsection{Event ambiguity processing}

The ambiguity solution is a combinatorial problem. The task of
the ambiguity processor is to decide about the association of (VD) hits and to
select the best tracks out of the set of mutually "excluded" candidates found
by the search algorithms.
The design of the DELPHI ambiguity processor \cite{fxsolv} was done in order
to find a balance between performance and CPU consumption.

The ambiguity processor maximises a "score" function for a given event.
The score of each track in the
solution is determined by the number of hits associated to the track and the
quality of the fit. A simple algorithm to resolve the event can start by
selecting the track with the highest score. The hits associated to that track
are removed from all other candidates. This implies refitting
the candidates from which a hit has been dropped. The list of candidates is therefore
changing in the course of the process. The process is iterated by selecting the
next best track until no more candidates are left over. This algorithm is
very fast, but any mistake at the beginning propagates through the rest of
the event analysis. Another algorithm, which
does not have this problem, would be to
create all possible lists of tracks, which contain no "exclusions" anymore,
in the same way as before. Here the list with the highest score would be
selected. This algorithm is limited by combinatorics, because the
number of lists increases very rapidly with the number of candidates.

The DELPHI ambiguity processor is a mixture of both algorithms.
It is a recursive algorithm, which in each step subdivides the event into
sets of "excluded" tracks to resolve them independently. For
each set all possible
lists of tracks are tried. One track after the other is taken out of the set
and each time the subset is resolved in the next
recursion level. For each recursion
the maximum possible score of the subset is calculated to truncate combinations
below the current maximum. A fall back solution is implemented, which uses
the simpler algorithm for a set in case it is not resolved after more than 2
minutes or the recursion depth is exceeding 9 levels.   

Additional protections are needed. Sub-tracks created during the
processing are rejected if they are only generated by splitting a long
track. A list of bad tracks is used to reject detector combinations of
poor quality or high risk of being fake.

The scoring function is tuned to optimise the track reconstruction efficiency
and the hit association purity at the same time. For each track a score of
100 is assigned, while a detector measurement associated to the track is
given a score between 1 and 20, depending on the quality of the
measurement, and a
logarithm of the $\chi^2$ probability of the track fit is added
to disfavour bad track candidates.  

The ambiguity processor is used three times in the track reconstruction
code. It is called for the first time to resolve the tracks including the
VD $R\phi$ hits, a second time to resolve the association of the $Rz$ hits
and finally to resolve ambiguities in the search dedicated to reconstruct
tracks before interactions with the material.

\subsection{Results of the new barrel track reconstruction package}

\begin{figure}[htb]
 \vspace{-1cm}
 \begin{center}
  \mbox{\hspace{-2.cm}\epsfig{file=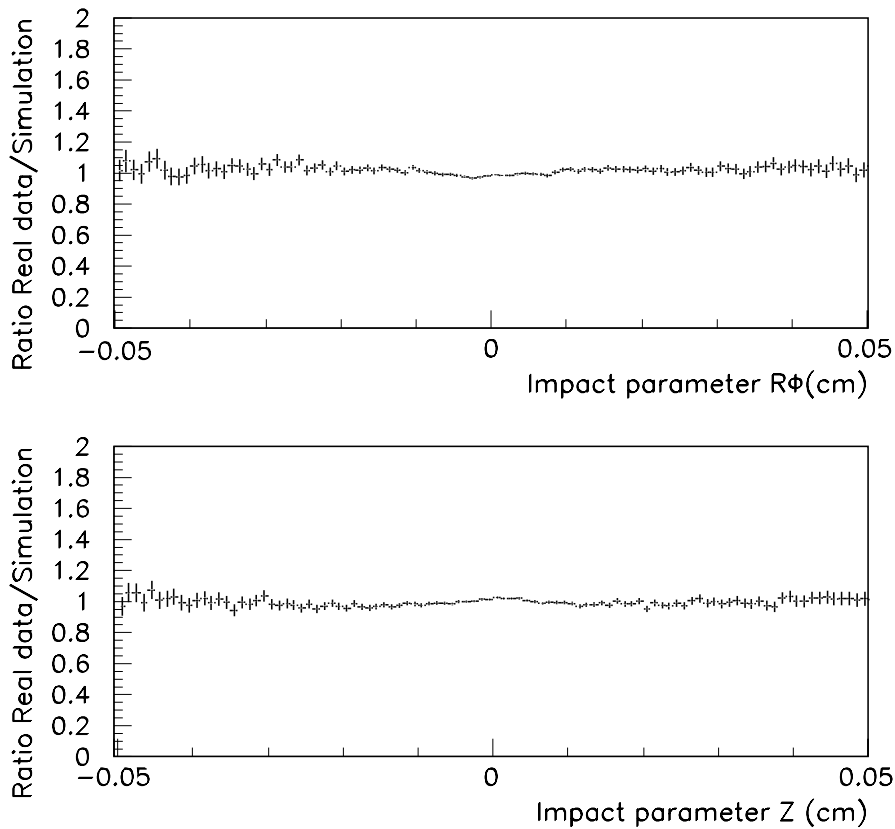,width=17cm}}
 \end{center}
 \vspace{-0.25cm}
 \caption[]{\label{imprat} The ratio of the impact parameter
in $R\phi$ and in $z$ measured in hadronic events for real data and
simulation. An excellent agreement in the impact parameter resolution
of reconstructed tracks in real data and simulation is found.}
\end{figure}

The new central tracking has been successfully used for the final
reprocessing of the full LEP 1 data set and for
the processing of the LEP 2 data.
An excellent reconstruction quality has been achieved and the precision of many DELPHI
physics results has been improved.

The impact parameter resolution for charged
tracks from hadronic $Z^0$ events has been measured \cite{VD-LEP2} to be :

\begin{eqnarray}
\sigma_{IP_{R\phi}} & = & \frac{71~ \mu m}{p\times\sin^{3/2}\theta}
                          \oplus 28 \mu m ~,\\
\sigma_{IP_{z}}     & = & \frac{75~ \mu m}{p\times\sin^{5/2}\theta}
                          \oplus 39 \mu m ~.
\end{eqnarray}

\noindent
In both cases the first term, which depends on the track polar angle $\theta$,
is the contribution due to multiple scattering, the
second term is the asymptotic value reflecting the measurement error.
The average miss distance at the interaction point between the two muons in
$Z \to \mu\mu$ events is measured to be 33 $\mu$m in $R\phi$ and 51.6 $\mu$m
in $Rz$ \cite{mumumiss}. In figure \ref{imprat} the ratio 
of the impact parameter distributions of reconstructed tracks from
hadronic $Z^0$ events in real data and simulation are shown separately for
$R\phi$ and $Rz$. The resolutions are correctly
described and the tails in the distributions due to wrong
association of VD hits to tracks are reproduced in the simulation.

\begin{figure}[htb]
 \vspace{-1cm}
 \begin{center}
  \epsfig{file=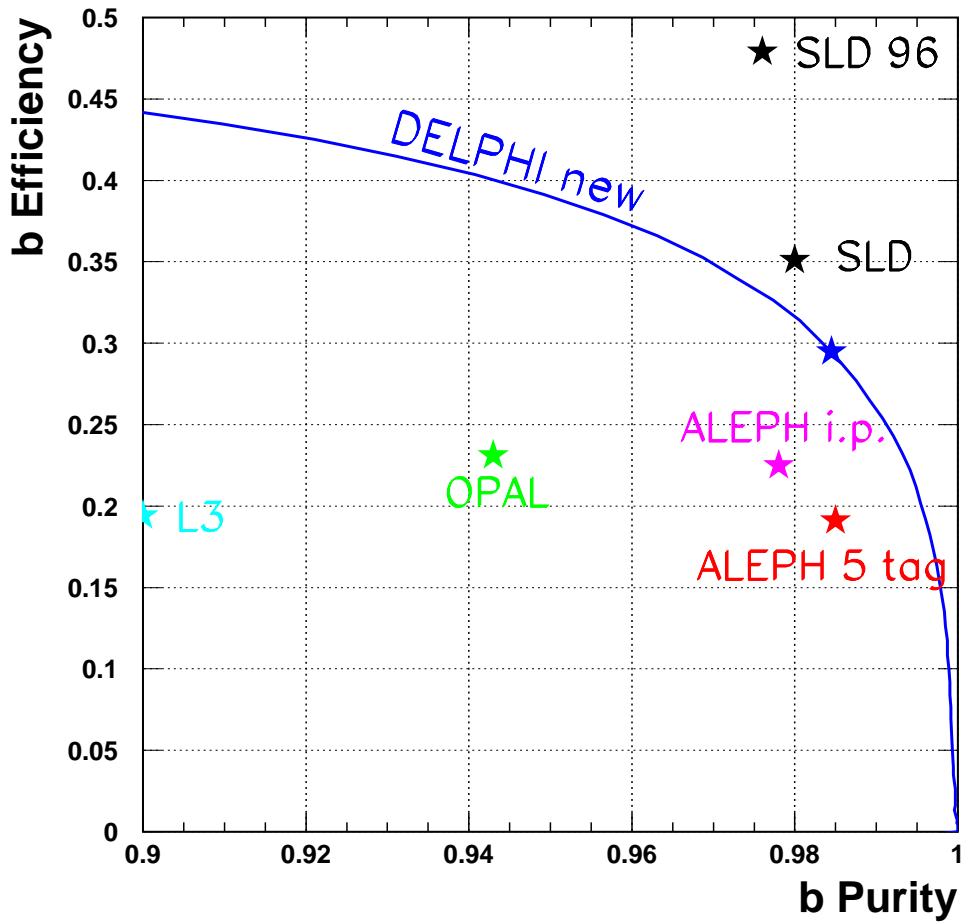,width=15cm}
 \end{center}
 \vspace{-1cm}
 \caption[]{\label{btag} A comparison of the $b$-tagging performance of the
LEP and SLC experiments. The working point for all experiments as well
as the efficiency vs purity contour for DELPHI determined by using the new
track reconstruction software is shown.}
\end{figure}

The most dramatic improvement due to the new reconstruction algorithms
is visible in the $b$-tagging performance
\cite{btagging}. The average number of tracks per hadronic $Z^0$ decay
used to determine the $b$ content increased from 9.5 to 14.3. Figure
\ref{btag} shows a comparison of the $b$-tagging efficiency as a function of
$b$ purity for different experiments. After the reprocessing DELPHI outperforms
all other LEP experiments. Only SLD has a better resolution due to its smaller
beam pipe and the smaller dimensions of the beam spot.

\begin{figure}[htb]
 \vspace{3cm}
 \begin{center}
  \mbox{\hspace{-1cm}\epsfig{file=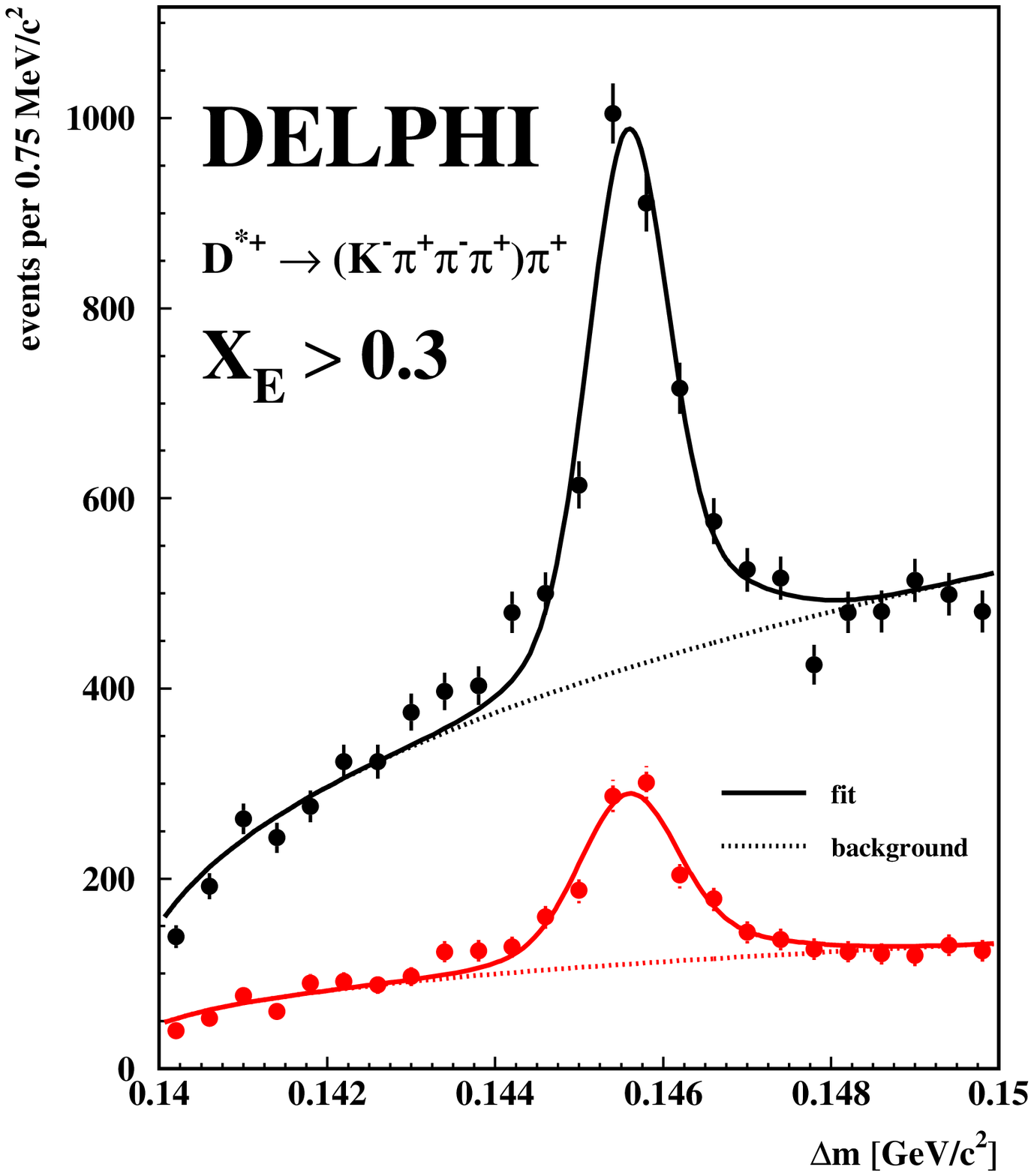,width=15cm}}
 \end{center}
 \vspace{-5cm}
 \caption[]{\label{dstar} The mass difference signal for the decay
$D^{*+} \to (K^-\pi^+\pi^-\pi^+)\pi^+$ using the 1994 data set. The smaller
signal is the result obtained using the old track reconstruction, the larger signal
shows the improvement introduced by the new track reconstruction on the same data
set.} 
\end{figure}

Figure \ref{dstar}
shows the mass difference between the $D^{*+}$ and the $D^0$ from
the decay $D^{*+} \to D^0 \pi^+$, where the $D^0$ decays into
$K^-\pi^+\pi^-\pi^+$. The signal is shown for both processings using the old
and the new track reconstruction code. In both cases the $D^{*+}$ decays are
reconstructed using the same analysis code and the same set of cuts. A gain
of a factor 2.5 in efficiency is observed for such complicated decay modes. 

\begin{figure}[htb]
 \begin{center}
  \epsfig{file=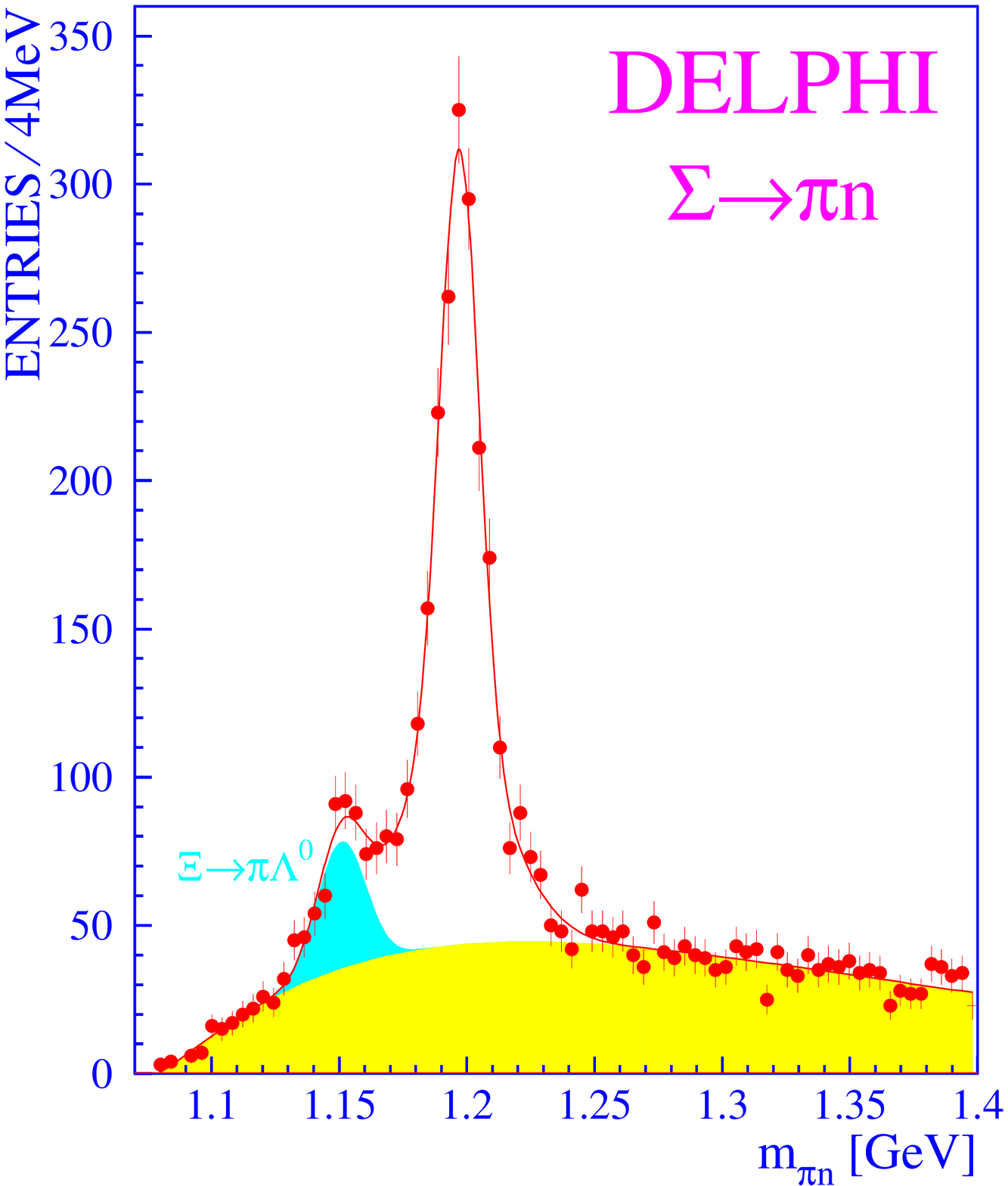,width=12cm}
 \end{center}
 \caption[]{\label{sigma} The $\Sigma \to \pi n$ mass spectrum reconstructed
using tracks associated to tagged decays in flight found by the vertex
searches. The $\Sigma$ tracks
have been measured in the VD. The decay pions are reconstructed in the
outer tracking system. A reflection from $\Xi \to \pi \Lambda^0$ decays is also
visible.}
\end{figure}

An example of an application \cite{DISweiser} of the secondary vertex search is
shown in figure \ref{sigma}. The mass signal of $\Sigma \to
\pi n$ decays is reconstructed from the $\Sigma$ tracks measured in the VD
and the tracks of the decay pions. The vertices have been found by the
search for decays in flight.

\begin{figure}[htb]
 \begin{center}
  \epsfig{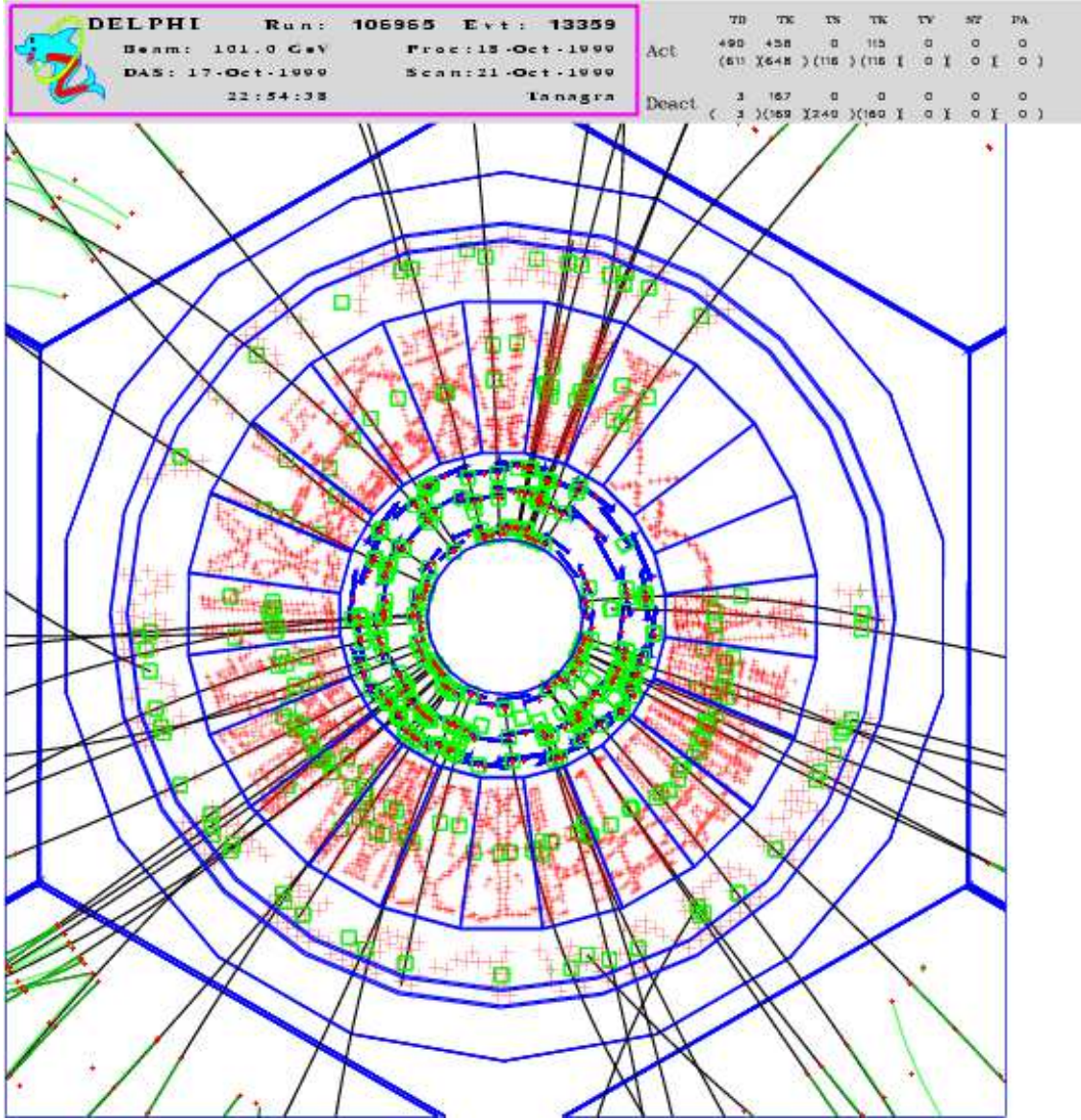}
 \end{center}
 \caption[]{\label{LEP2} A reconstructed high multiplicity hadronic event taken
at a centre-of-mass energy of 202 GeV in 1999. Shown is an $R\phi$ view of the
central part of the detector starting from the VD layers, the ID jet chamber
and straw tubes up to the inner radius of the TPC.}
\end{figure}

Figure \ref{LEP2} shows an example of a reconstructed high multiplicity
multi-jet event taken at an centre-of-mass energy of 202 GeV in the year
1999. No significant degradation of the $b$-tagging resolution has been
observed for such complicated events.

\section{VFT standalone tracking and the acceptance in the forward region}

In contrast to the barrel tracking the track reconstruction strategy in the
forward region is completely different. In this region the
VFT is needed to improve
the acceptance for charged particles. Standalone track reconstruction
in the VFT is mandatory to measure the tracks before most of the
particles shower in the material of the end rings of the barrel detectors.

\subsection{VFT standalone track reconstruction}

The VFT (see figure \ref{VDfig}) consists of two layers of pixel detectors and
two layers of back-to-back mini strip detectors.
It gives on average 2 to 3 space points per track in
the polar angle range from $21^\circ$ to $10.5^\circ$. An important aspect of
the pixel detector is the low random noise rate of $0.5 \times 10^{-6}$
after an offline suppression of 0.3 \% of systematically noisy pixels
\cite{VD-LEP2}. This ensures a small
rate of fake hits and consequently a good purity in the reconstruction.
The reconstructed clusters in both views of a back-to-back mini strip module
are combined to obtain 3 dimensional hit information.

The standalone tracking \cite{VD-LEP2} is done in 3 steps. First
all combinations
of 3 layers are tested and the track parameters are determined using a
helix fit. One requires the reconstructed track elements to point towards the
primary interaction region. The primary interaction region has a dimension
of 0.77 $c$m in $z$ and of 150 and 10 $\mu$m in $x$ and $y$,
respectively \cite{murray}. In the next step the track finding efficiency
is improved using 
all left over two hit combinations in both pixel layers. The
point of the average primary interaction position is
added to the combinations to determine the track
parameters. Finally a similar strategy
is used for combinations of space points in both mini strip layers. A $2^\circ$
stereo angle and flipping of the module orientation leads to a relative
angle of
$4^\circ$ between the strips in the same projection of both layers. This
relative angle is used to remove fake
combinations of mirror images which do no longer
point towards the primary interaction region. The result of the standalone
tracking as well as all hits are then used in the forward search
to reconstruct the full track.

\subsection{The VFT in the forward track search}

\begin{figure}[t]
 \vspace{1cm}
 \begin{center}
  \mbox{\hspace{-1cm}\epsfig{file=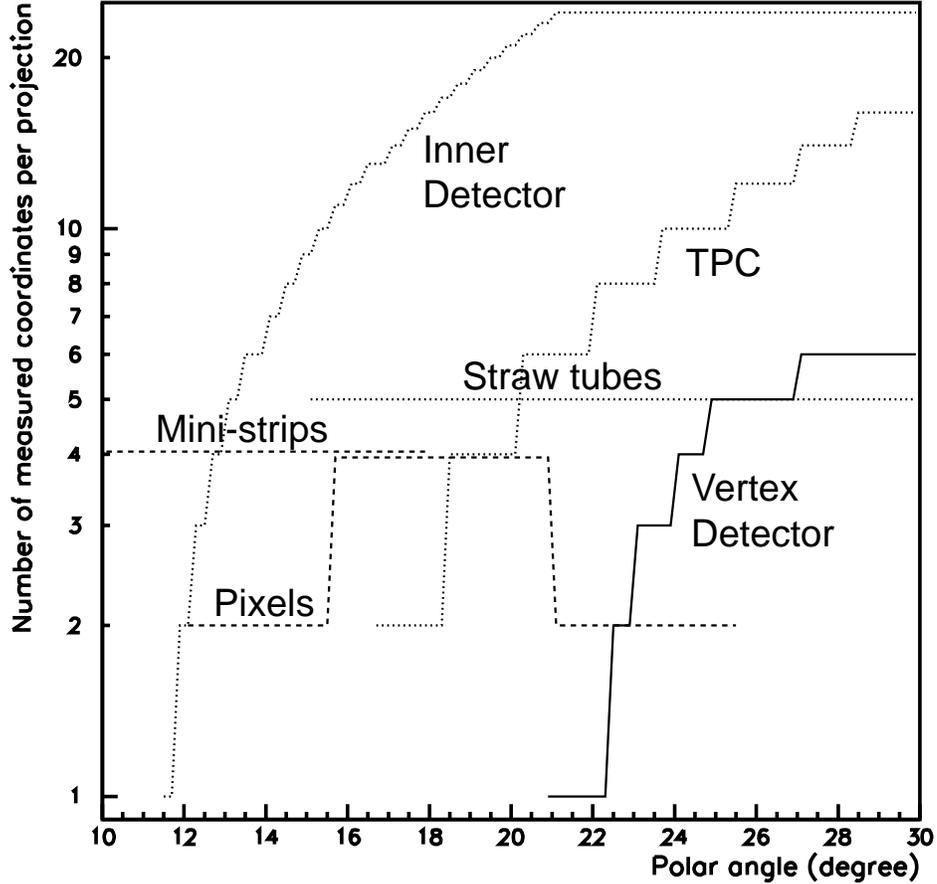,width=11cm}}
 \end{center}
 \vspace{4cm}
 \caption[]{\label{forward} The number of measured points as a function of the
polar angle for the different
tracking detectors in the forward region of DELPHI. The forward chambers are
not shown.}
\end{figure}

The forward track reconstruction \cite{tksf} is limited by the material of
the end rings
of the barrel detectors, which amounts to 1.4 radiation length in front of the
electromagnetic calorimeter. Furthermore tracks are dropping out of the 
acceptance of the central detectors with decreasing polar angle
as it is shown in figure \ref{forward}, where
the number of measurements can be seen as a function of polar angle for all
detectors but the forward chambers. The latter contribute 18 additional points
over the full range, but they operate behind 
the material of the end rings of the barrel detectors.

The VFT, which is close to the interaction region, is the
basis of the track forward finding. The
track search algorithm uses as starting points as many different
seeds as possible. These seeds are either
VFT tracks from the
standalone tracking, combinations of VFT hits with ID jet chamber hits or
other detector combinations like VD and ID jet chamber or TPC. In total 12
different combinations are tried. Starting from each of the seeds 
a simple road search is done to look for possible hits in the other detectors
to be associated to the track candidate. On the resulting list of all possible
hits from all detectors a search is done for track combinations
including the maximum number of detectors. The Kalman Filter track
fit with its outlier logic serves as the final filter to select good
candidates, which are fed into the global event solution to resolve
ambiguities. At this stage of the reconstruction no tracks 
measured only in the VFT and in the VFT and the ID
jet chamber are considered. A dedicated search is done afterwards to 
reconstruct these tracks out of the remaining
tracks found by the VFT standalone track reconstruction.

\subsection{Results of the new forward tracking}

\begin{figure}[htb]
 \vspace{-4cm}
 \begin{center}
  \epsfig{file=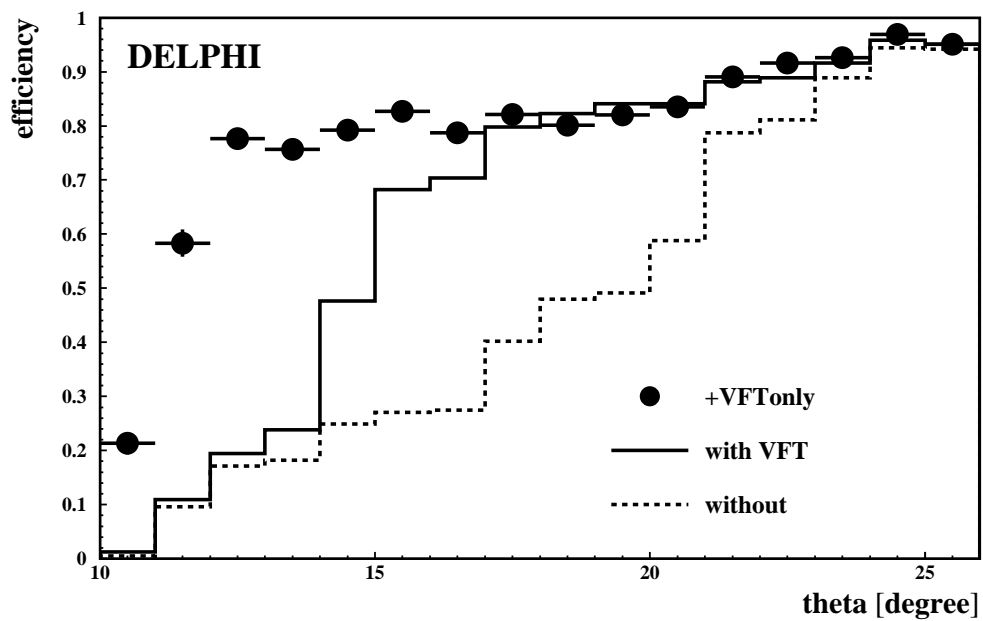,width=13.5cm}
 \end{center}
 \caption[]{\label{effi} The efficiency to reconstruct primary tracks in simulated
hadronic $Z^0$ events as a function of the polar angle. Shown are the results
with and without the VFT in the track reconstruction as well as the additional
gain due to the tracks reconstructed only in the VFT.}
\end{figure}

\begin{figure}[htb]
 \begin{center}
  \epsfig{file=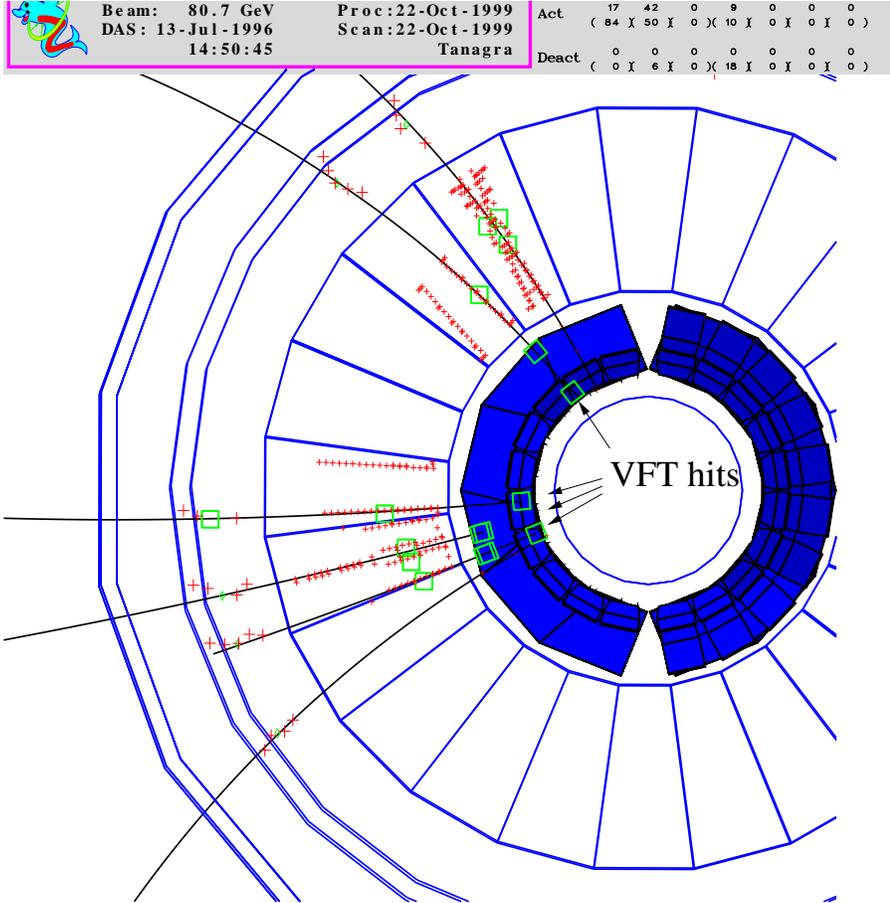,width=12cm}
 \end{center}
 \caption[]{\label{gamgam} A $\gamma\gamma$ event tagged by an electron seen
in the Small Angle Tile Calorimeter. Shown is an $R\phi$ view of the VFT,
the ID and the TPC. Four out of six tracks in the event are reconstructed
using the hits measured in the VFT.} 
\end{figure}

Figure \ref{effi} shows the improvement of the track finding efficiency for
primary tracks from the interaction region as
a function of the polar angle $\theta$ after including the VFT into the
forward track reconstruction. A clear gain is visible
down to $14^\circ$ for tracks reconstructed using the VFT. Below $13^\circ$ the
ID jet chamber drops out of the tracking and the VFT standalone tracks
are used to extend the efficiency plateau down to $11^\circ$. An example of a
$\gamma\gamma$ event is shown in figure \ref{gamgam}. In the event one electron
is measured in the Small Angle Tile Calorimeter. Most of the tracks are
reconstructed using the VFT hit information.

\section{Conclusion}

The DELPHI Silicon Tracker is used successfully in two
different tracking situations. In the barrel region the VD
provides precision tracking information. Optimised algorithms were developed to
reconstruct the charged tracks starting from the VD hits. The result of this
new reconstruction software is an excellent
data quality which gives DELPHI the best $b$-tagging performance of all LEP
experiments. In the forward region different reconstruction strategies based
on the VFT are used to significantly improve the track reconstruction
efficiency down to a polar angle of $10.5^\circ$.


\section*{Acknowledgements}

I would like to thank P. Bruckman, K. \"Osterberg, M.E. Pol and C. Weiser for
providing me with material for the presentation. I also like to thank 
K. \"Osterberg for reading the article and sending constructive comments.


\end{document}